# Reductive Arylation of Graphene: Insights into a Reversible Carbon Allotrope Functionalization Reaction

Philipp Vecera[1,2], Konstantin Edelthalhammer[2], Frank Hauke[1] and Andreas Hirsch[*,1,2]

[1]Institute of Advanced Materials and Processes (ZMP), Friedrich-Alexander-Universität Erlangen-Nürnberg (FAU), Fürth, Dr.-Mack-Straße 81, 90762 Fürth

[2]Department of Chemistry and Pharmacy, Friedrich-Alexander-Universität Erlangen-Nürnberg (FAU), Henkestraße 42, 91054 Erlangen

*e-mail andreas.hirsch@fau.de, Phone: +49 9131 85 22537, Fax: +49 131 85 26864

The covalent functionalization of graphene represents a main topic in the growing field of nano materials. The reductive exfoliation of graphite with concomitant functionalization of the respective graphenide intermediates provides a promising approach towards functional graphene derivatives. In this article, we present new insights into the reductive arylation of graphene. Graphite intercalation compounds (GICs) with varying stoichiometries have been used as starting materials. Based on the spectroscopic data obtained by scanning Raman microscopy (SRM) and thermogravimetric analysis coupled to mass spectrometry (TG/MS), a clear correlation between the amount of negative charges - present in the GIC - and the degree of functionalization in the final product could be found. Furthermore, the detailed analysis of the thermal defunctionalization process provided deeper insights into the covalent addend binding.

**1 Introduction** Driven by their remarkable properties, synthetic carbon allotropes (SCAs) and in particular the 2-dimensional representative - graphene - have inspired the field of research of novel materials throughout the last decade.[1] The development of cost efficient production techniques of graphene - in sufficient quantities and qualities, suitable for potential applications - still remains challenging. A solution to this problem may be provided by wet chemical approaches on the basis of the cheap starting material graphite, where simultaneously by means of direct covalent functionalization novel materials properties are accessible.[2-4]

Since the pioneering work of Pénicaud *et al.*[5] graphite intercalation compounds (GICs) have been exploited as highly activated species for a subsequent covalent functionalization of the individualized carbon layers.[6] The trapping of the negatively charged intermediates – termed graphenides – by suitable electrophiles in inert solvents has been developed to a well-established approach for the synthesis of graphene derivatives, with functional entities covalently attached to an intact carbon framework.[7-9]. In contrast to oxidative routes applied for the production of graphene oxide and its respective derivatives,[10, 11] reductive functionalization protocols inherently bear the advantage for a complete reversible thermal de-functionalization with a recovery of a defect-free graphene lattice. Along the development of efficient functionalization protocols, tools for the characterization of the reaction products have continuously been adjusted and improved. On the basis of high-end analytical techniques, like Raman spectroscopy[12] and thermogravimetric analysis coupled to mass spectrometry (TG/MS), we are now able to quantify the density of functional groups bound to the respective SCA lattice.[13, 14]

In this manuscript we used the reductive arylation of graphite, with phenyl iodide as trapping electrophile, as a benchmark reaction (Figure 1) for the investigation of the influence of the potassium to carbon ratio (K:C = 1:4 – K:C = 1:48) on the degree of functionalization in the final product $G_{(K:C)}Ph$. As we have shown recently for carbon nanotubes, the variation of the amount of potassium within the reductive activation step is one fundamental parameter to adjust the degree of functionalization in the respective CNT derivatives.[15]

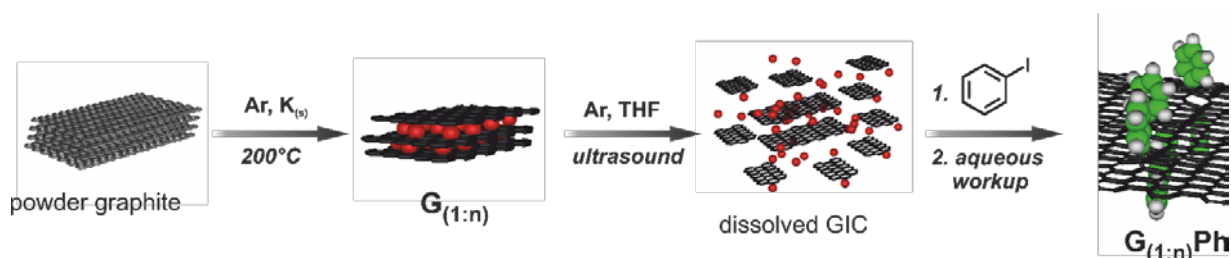

**Figure 1:** Reaction scheme for the production of $G_{(1:n)}Ph$ by the use of graphite as starting material for the reductive arylation of graphenides in THF.

For a better understanding of the chemical functionalization sequence, the reaction progress starting from the initially prepared GIC $G_{(1:8)}$ [$C_8K$] was monitored *in situ* and under inert gas conditions by Raman spectroscopy. Furthermore, we present novel insights into the thermal annealing of chemically functionalized graphene derivatives. By a correlation of the data obtained by thermogravimetric analysis coupled to mass spectrometry (TG/MS) with the information acquired by temperature dependent Raman spectroscopy we were able to show that the approach of reductive functionalization of GICs is completely reversible and that the defect-free honeycomb lattice can easily be restored by a bulk production approach.

**2 Experimental** The complete chemical functionalization sequence, excluding the final aqueous work-up, was carried out in a glove box under inert gas conditions (argon atmosphere, <0.1 ppm $O_2$, <1 ppm $H_2O$). As starting material a spherical powder graphite [Future Carbon© SGN18] (see also Figure S1) was used. By stoichiometric melting with potassium, the respective GIC – $G_{(1:4)}$, $G_{(1:8)}$, $G_{(1:10)}$, $G_{(1:24)}$, $G_{(1:48)}$ – was synthesized and dissolved in absolute THF by short ultrasonication pulses (tip sonicator). The intermediately formed graphenides were subsequently trapped by the addition of two equivalents of phenyl iodide. The reaction was quenched by the addition of water and diluted HCl (neutralization to pH =7). The functionalized material $G_{(1:n)}Ph$ was isolated by phase separation of a water/THF mixture with subsequent filtration and washing. Analysis of the bulk material was carried out by Scanning Raman Microscopy (SRM) and thermogravimetric analysis coupled to mass spectrometry (TG/MS).

**3 Results and Discussion** In order to gain a statistically significant bulk information about the success of the reductive arylation, SRM on large sample areas (100 x 100 µm) has been carried out. In particular, for all samples the $I_{(D/G)}$ (intensity ratios of the D- and G-modes) values of the individual Raman spectra are plotted with respect to their frequency and the respective distribution function was determined (Figure S3) – mean Raman spectra of compounds **G$_{(1:n)}$** are presented in Figure 2 (top). In general, the D-mode intensity is a measure for the amount of $sp^3$-defects introduced by covalent addend binding.[12] In Figure 2 (bottom), the successful arylation of the reductively charged graphite intercalation compounds is exemplarily demonstrated for **G$_{(1:8)}$Ph** – mean $I_{(D/G)}$ ratio of 1.4.

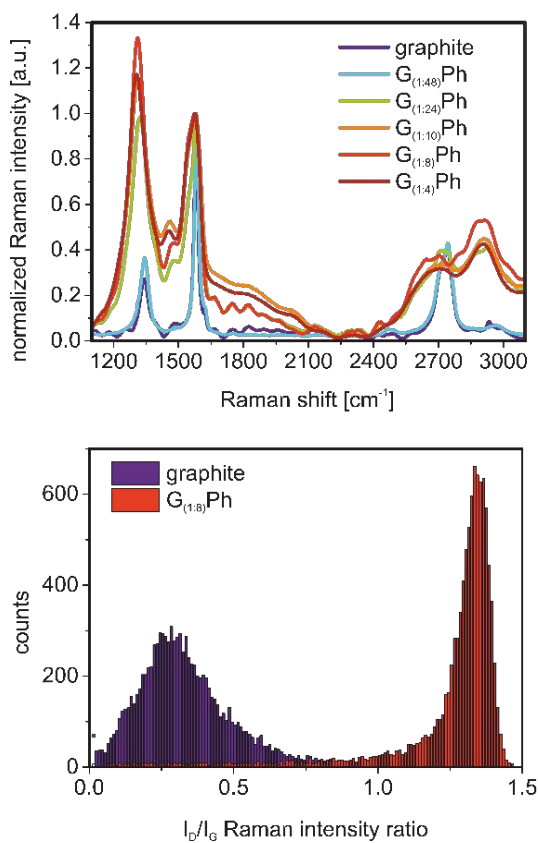

**Figure 2**: Statistical Raman spectroscopic analysis of the reductively arylated samples. Top: Averaged (10.000 single point spectra) Raman spectra obtained for samples with varying potassium to carbon ratios. Bottom: Histogram of the SGN18 starting material in comparison to **G$_{(1:8)}$Ph**.

Based on the bulk material Raman analysis no profound conclusions on the efficiency of the exfoliation process and the average number of layers in the final reaction product can been drawn. Therefore, the supernatant of the **G$_{(1:8)}$Ph** dispersion has been drop-casted on silicon/SiO$_2$ (300 nm oxide layer) substrates.[16] Here, separated flakes (mean diameter 1 – 5 µm) can be identified by optical microscopy and statistically be analyzed by SRM – Figure S4. The reductive exfoliation with subsequent arylation predominantly yields highly functionalized few layer material. Besides that, also spots with functionalized bilayer graphene with a sharp and symmetric 2D-mode (2718 cm$^{-1}$ - full-width at half-maximum ($\Gamma_{2D}$) of 49 cm$^{-1}$) and functionalized mono-layer graphene ($\Gamma_{2D}$ = 29 cm$^{-1}$) can be detected. For the latter ones, the spectral data is in line with the characteristic signatures of the "low density of defect area" described by Lucchese *et al.*[17] and Cançado *et al.*[18] Hence, based on these results a degree of functionalization of approx. 0.02% (per lattice C-Atom) can be estimated for these systems.[14] Nevertheless, the main part of the material remained in the residue and based on the bulk Raman analysis, a higher degree of functionalization can be attributed to this fraction.

For a profound insight into the type of material generated by the reductive exfoliation of **G$_{(1:8)}$**, we analyzed these graphenide flakes by means of *in situ* Raman spectroscopy. Here, the intermediates have been isolated under inert gas conditions by filtration and the highly air sensitive material has been transferred in a special air-tight setup into the Raman spectrometer.[19] The charged powder exhibits a single Raman signal at 1607 cm$^{-1}$ (shifted G-mode, $\Gamma$ of 20 cm$^{-1}$) characteristic for doped and decoupled graphene sheets (Figure 3).[12] Remarkably, no pronounced D-mode intensity is found for this material, which speaks for the fact that flakes with sufficient surface area are generated during the exfoliation step, as no significant D-mode contribution of rim carbon atoms is detected. This directly leads to the conclusion that the pronounced D-mode intensity increase detected after the addition of phenyl iodide can be attributed to a direct covalent addition of the phenyl moieties to the respective framework carbon atoms of the individual sheets.

We have shown recently that the variation of the amount of potassium within the reductive activation step of carbon nanotubes is one fundamental parameter with direct influence on the obtained degree of functionalization.[15] In order to investigate the respective relation for graphite as carbon allotrope starting material, the potassium:carbon stoichiometry was varied as follows: 1:48, 1:24, 1:10, 1:8, 1:4 – for details see ESI. In all cases the amount of the trapping electrophile phenyl iodide was kept constant (two-fold molar excess with respect to graphite carbon lattice atoms). The Raman spectroscopic and thermogravimetric data obtained for the final derivatization products are summarized in Figure 4. Obviously, an increasing amount of potassium used for the GIC formation directly correlates with an increase of the $I_{(D/G)}$ ratio and, moreover, with the mass loss detected by TG analysis. This can easily be understood in terms of the underlying reaction mechanism.[20] Here, on the basis of an electron transfer reaction originating from the negatively charged intermediates towards the phenyl iodide trapping reagent, an aryl radical is formed in close proximity to the carbon scaffold.

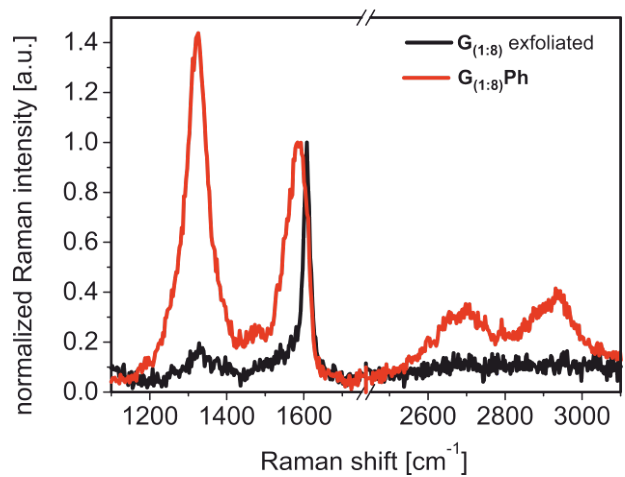

**Figure 3:** *In situ* Raman analysis of the graphenide intermediates in comparison to the arylated product **G$_{(1:8)}$Ph**. Black: Exfoliated **G$_{(1:8)}$** (powder isolated under inert gas conditions from THF dispersions) prior to the addition of phenyl iodide. Red: Median spectrum of **G$_{(1:8)}$** based on the Voigt distribution of the functionalized bulk material.

Subsequently, this carbon centered radical attacks the extended aromatic π-system and the phenyl ring is covalently bound to the carbon framework. Therefore, an increase of the charge density (increasing potassium amount) directly correlates with the degree of functionalization. This is fully in line with the results obtained for carbon nanotubes.[15] Consistently, the highest $I_{(D/G)}$ value is found for C$_8$K (**G$_{(1:8)}$**) – the 1$^{st}$ stage intercalation compound of graphite. This stoichiometry represents the highest possible potassium doping level and directly related to that, the highest possible charging state of the individual graphene layers. Interestingly, an excess of potassium present in the reaction medium does not lead to higher degrees of functionalization. This observation can be explain by the fact, that aryl radicals formed in close proximity to these excess potassium atoms can not diffuse to the dispersed graphene sheets, but recombine in a dimerization step. On the other hand, the amount of phenyl iodide has no influence on the obtained degree of functionalization as long as an excess of trapping electrophile is used in respect to the present charge density in the graphenide intermediates. This rationale is corroborated by a reference experiment, where the optimum K:C ratio - found in the 1$^{st}$ stage GIC **G$_{(1:8)}$** [C$_8$K] - was used. Here, the amount of phenyl iodide was varied in the range of 1.6 – 80 equivalents with respect to the amount of negative charges present in **G$_{(1:8)}$**. The respective Raman data (Figure S5) clearly demonstrate that for all phenyl iodide concentrations, the same degree of functionalization is reached in the final derivatization product – TG data is presented in Figure S7. This nicely underlines that the amount of reductively applied charges is the predominant parameter, responsible for the overall degree of functionalization in this type of reductive arylation approaches of GICs.

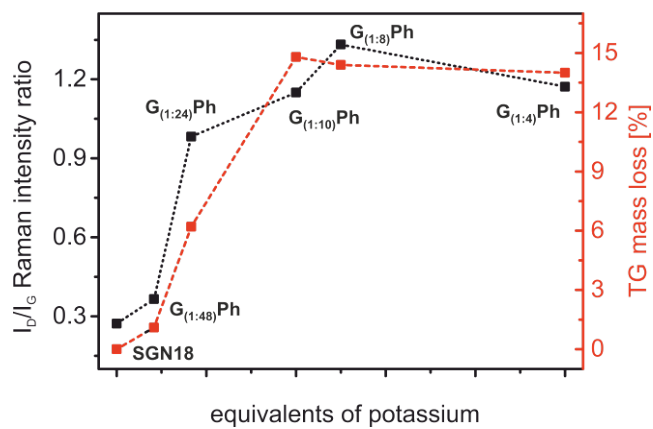

**Figure 4:** Correlation between equivalents of potassium used for the reductive exfoliation and Raman $I_{(D/G)}$ intensity ratio as well as sample mass loss detected by TGA. Maximum degree of functionalization is reached for the 1$^{st}$ stage intercalation compound $C_8K$.

**3.2 Thermal Defunctionalization** In general, Raman spectroscopy provides a direct insight into the introduction of sp$^3$-defects but does not yield any information about the chemical nature of the bound addends. For this purpose, the reductively arylated derivative **G$_{(1:8)}$Ph** has been analyzed by means of thermogravimetric analysis coupled to mass spectrometry (TG/MS) – (Figure 5, for TG profiles see Figures S8). Here, the sample is heated under an inert gas atmosphere and the thermal detachment of the phenyl moieties ($^m/_z$ 77) is constantly monitored by mass spectrometry and can be correlated to the mass losses observed in the TGA. The two main steps of mass loss (250 °C, 500 °C) detected by TG nicely correlated with the maximum ion current of the phenyl ring fragments. At first glance, the mass loss in the lower temperature region may be assigned to the detachment of physisorbed species whereas the mass loss at higher temperatures may be attributed to the detachment of chemisorbed (covalently bound) moieties.

This would be in clear contrast to the temperature dependent Raman measurements (TDRS) which have been carried out for **G$_{(1:8)}$Ph** (Figure 6 – for full set of data see Figure S8). Here, a constant decrease of the $I_{(D/G)}$ ratio is detected over the whole temperature range, which also nicely correlates with the respective TG profile (Figure 5). As the detachment of physisorbed species would not result in a rehybridization of sp$^3$ lattice atoms, the temperature dependent Raman data is indicative for a thermal cleavage of covalently bound entities, even in the low-temperature regions.

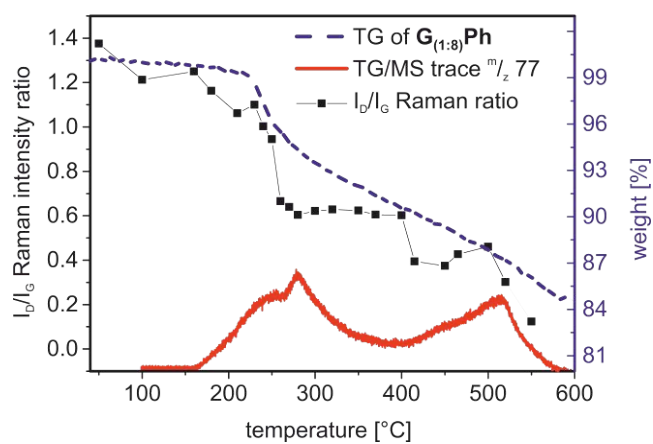

**Figure 5:** Mass loss and respective Raman $I_{(D/G)}$ intensity ratio (extracted from the temperature dependent Raman spectra) of $G_{(1:8)}Ph$ upon heating under inert gas atmosphere. Based on the detected ion current for $^{m}/_{z}$ 77 ($C_6H_5$) two main regions for functional group cleavage can be extracted: 250 °C and 500 °C.

This observation may be explained by the existence of different binding sites: rim, edge near positions, in-plane position. This assumption is currently under investigation in our lab. Nevertheless, the combination of the two complementary analytical techniques – TG/MS and TDRS - yields new basic insights into the covalent chemistry of graphene.

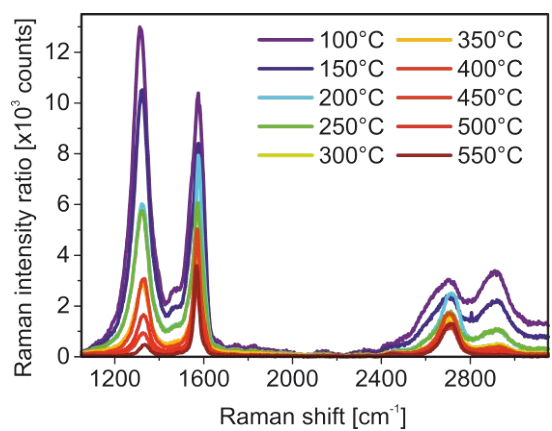

**Figure 6:** Temperature dependent Raman spectroscopy (TDRS) of $G_{(1:8)}Ph$ in the temperature region between 100 °C and 550 °C.

**4 Conclusion** In this manuscript, we presented new basic insights into the wet chemical exfoliation of graphite and the concomitant reductive arylation of the graphenide intermediates. We have demonstrated, that the amount of negative charges - present in the respective graphite intercalation compounds $G_{(1:n)}$ - clearly determines the degree of functionalization in the final product. Moreover, the degree of functionalization can be maximized when $G_{(1:8)}$ [$C_8K$] is used as starting material for the reductive functionalization approach. Our study corroborates that this type of GIC functionalization approach is based on a radical reaction where the negatively charged carbon layers act as electron donor components. The arylated graphene derivatives have been analyzed in detail by scanning Raman microscopy (SRM), temperature dependent Raman spectroscopy (TDRS) and thermogravimetric analysis coupled to mass spectrometry (TG/MS). Based on the complementary data set, a deeper understanding of the thermal detachment of covalently bound moieties was obtained.

**Acknowledgements** The authors thank the Deutsche Forschungsgemeinschaft (DFG-SFB 953 "Synthetic Carbon Allotropes", Project A1) and the Graduate School Molecular Science (GSMS) for financial support. The research leading to these results has received funding from the European Union Seventh Framework Programme under grant agreement n°604391 Graphene Flagship.

**References**

[1] A. K. Geim, K. S. Novoselov, Nat. Mater. **6**, 183-191 (2007).
[2] J.E. Johns, M.C. Hersam, Acc. Chem. Res. **46**, 77-86 (2013).
[3] C. K. Chua, M. Pumera, Chem. Soc. Rev. **42**, 3222-3233 (2013).
[4] L. Rodríguez-Pérez, M. Á. Herranz, N. Martín, Chem. Commun. **49**, 3721-3735 (2013).
[5] A. Pénicaud, C. Drummond, Acc. Chem. Res. **46,** 129-137 (2013).
[6] A. Hirsch, J. M. Englert, F. Hauke, Acc. Chem. Res. **46**, 87-96 (2013).
[7] J. M. Englert, C. Dotzer, G. Yang, M. Schmid, C. Papp, J. M. Gottfried, H.-P. Steinrück, E. Spiecker, F. Hauke, A. Hirsch, Nat. Chem. **3**, 279-286 (2011).
[8] J. M. Englert, K. C. Knirsch, C. Dotzer, B. Butz, F. Hauke, E. Spiecker, A. Hirsch, Chem. Commun. **48**, 5025-5027 (2012).
[9] K. C. Knirsch, J. M. Englert, C. Dotzer, F. Hauke, A. Hirsch, Chem. Commun. **49**, 10811-10813 (2013).
[10] Y. Zhu, D. K. James, J. M. Tour, Adv. Mater. **24**, 4924-4955 (2012).
[11] D. R. Dreyer, S. Park, C. W. Bielawski, R. S. Ruoff, Chem. Soc. Rev. **39**, 228-240 (2010).
[12] L. M. Malard, M. A. Pimenta, G. Dresselhaus, M. S. Dresselhaus, Phys. Reports **473**, 51-87 (2009).
[13] F. Hof, S. Bosch, J. M. Englert, F. Hauke, A. Hirsch, Angew. Chem. Int. Ed. **51**, 11727-11730 (2012).
[14] J. M. Englert, P. Vecera, K. C. Knirsch, R. A. Schäfer, F. Hauke, A. Hirsch, ACS Nano **6**, 5472–5482 (2013).
[15] F. Hof, S. Bosch, S. Eigler, F. Hauke, A. Hirsch, J. Am. Chem. Soc. **135**, 18385–18395 (2013).
[16] D. S. L. Abergel, A. Russell, V.I . Falko, Appl. Phys. Lett. **91**, 063125 (2007).
[17] M. M. Lucchese, F. Stavale, E. H. M. Ferreira, C. Vilani, M. V. O. Moutinho, R. B. Capaz, C. A. Achete, A. Jorio, Carbon **48**, 1592-1597 (2010).
[18] L. G. Cançado, A. Jorio, E. H. M. Ferreira, F. Stavale, C. A. Achete, R. B. Capaz, M. V. O. Moutinho, A. Lombardo, T. Kulmala, A. C. Ferrari, Nano Lett. **11**, 3190–3196 (2011).
[19] A. M. Dimiev, G. Ceriotti, N. Behabtu, D. Zakhidov, M. Pasquali, R. Saito, J. M. Tour, ACS Nano **7**, 2773-2780 (2013).
[20] J. Chattopadhyay, S. Chakraborty, A. Mukherjee, R. Wang, P. S. Engel, W. E. Billups, J. Phys. Chem. C **111**, 17928-17932 (2007).

# *Supporting Information*

## Reductive Arylation of Graphene: Insights into a Reversible Carbon Allotrope Functionalization Reaction

**Philipp Vecera[1,2], Konstantin Edelthalhammer[2], Frank Hauke[1] and Andreas Hirsch**[\*,1,2]

[1] Institute of Advanced Materials and Processes (ZMP), Friedrich-Alexander-Universität Erlangen-Nürnberg (FAU), Fürth, Dr.-Mack-Straße 81, 90762 Fürth
[2] Department of Chemistry and Pharmacy, Friedrich-Alexander-Universität Erlangen-Nürnberg (FAU), Henkestraße 42, 91054 Erlangen

## 1. Materials

As starting material a spherical graphite SGN18 (Future Carbon, Germany), a synthetic graphite (99.99 %C, <0.01% ash) with a comparatively small mean grain size of 18 µm (Figure S1), a high specific surface area of 6.2 m$^2$/g and a resistivity of 0.001 Ωcm was chosen. As has been shown recently[1] SGN18 provides the ideal basis for an efficient generation of individualized graphene flakes by a reductive exfoliation approach. Furthermore, this material yields functionalized graphene derivatives with a higher degree of functionalization in comparison to other types of graphite, like natural flake or powder graphite.

Chemicals and solvents were purchased from Sigma Aldrich Co. (Germany) and were used as-received if not stated otherwise.

[1] K.C. Knirsch, J.M. Englert, C. Dotzer, F. Hauke, A. Hirsch, Chem. Commun., **49**, 10811-10813 (2013).

[*] e-mail andreas.hirsch@fau.de, Phone: +49 9131 85 22537, Fax: +49 131 85 26864

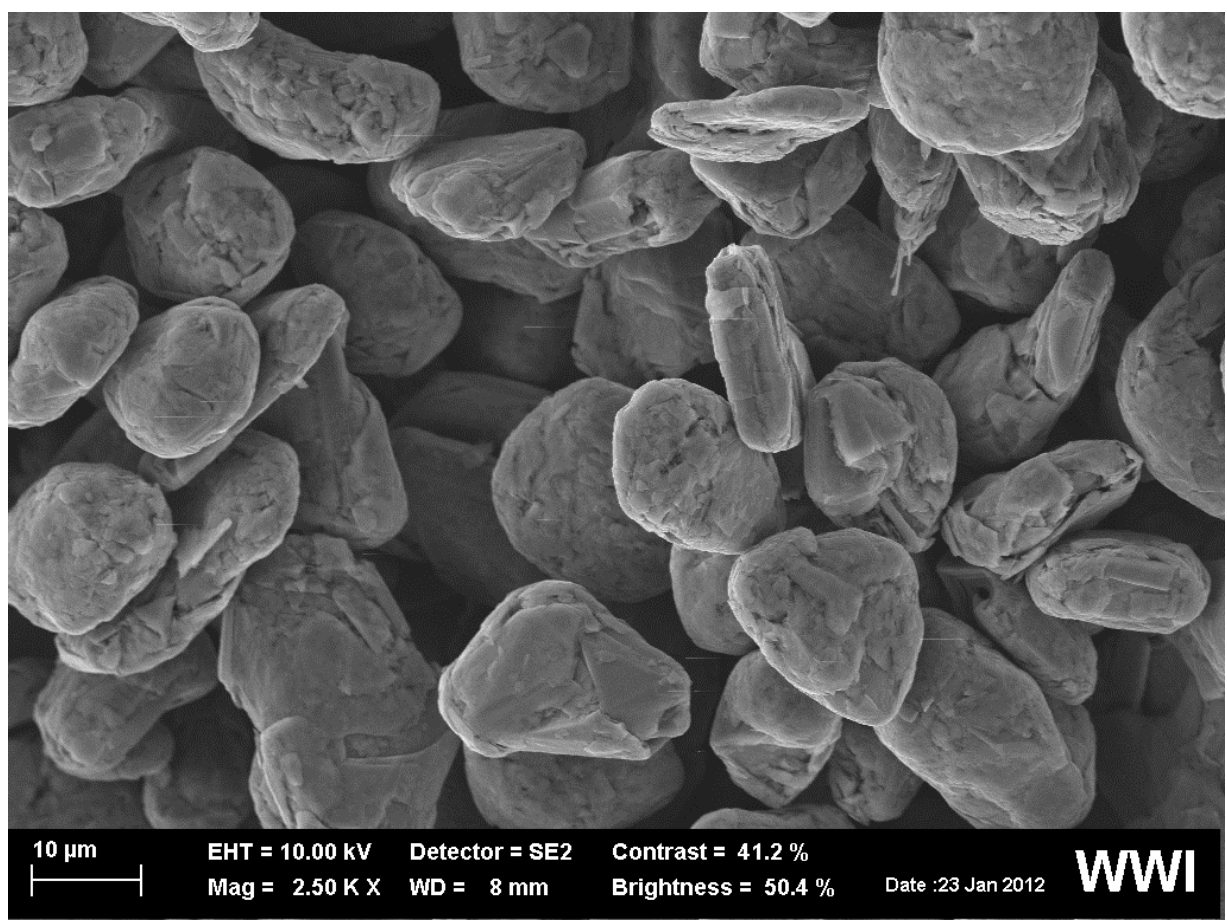

**Figure S1:** SEM image of the graphitic starting material SGN18 – Magnification 2500x.

THF was distilled three times in an argon inert gas atmosphere in order to remove residual water: a) over $CaH_2$, b) over sodium and c) over sodium-potassium alloy. Residual traces of oxygen were removed by pump freeze treatment (3 iterative steps). $THF_{(abs)}$ was used for all reactions.

Phenyl iodide was distilled in an argon inert gas atmosphere. Residual traces of oxygen were removed by pump freeze treatment (1 step).

## 2. Equipment and Characterization

*Glove Box:* Sample functionalization was carried out in an argon filled Labmaster sp glove box (MBraun), equipped with a gas filter to remove solvents and an argon cooling systems, with an oxygen content <0.1 ppm and a water content <0.1 ppm.

***Raman Spectroscopy:*** Raman spectroscopic characterization was carried out on a HoribaLabRAM Aramis confocal Raman microscope ($\lambda_{exc}$ : 532) with a laser spot size of about 1 μm (Olympus LMPlanFl 50x, NA 0.50) in backscattering geometry. The incident laser power was kept as low as possible to avoid structural sample damage: 1.35 mW (532 nm). Spectra were recorded with a CCD array at -70 °C – grating: 600 grooves. For a statistical significant bulk analysis, Raman spectra were obtained from a 100 x 100 μm area with 1 μm step size in SWIFT mode for low integration times. Calibration in frequency was carried out with a HOPG crystal as reference. Sample movement was carried out by an automated XY-scanning table.

Temperature depending Raman measurements were performed in a Linkam stage THMS 600, equipped with a liquid nitrogen pump TMS94 for temperature stabilization under a constant flow of nitrogen. The measurements were directly carried out on powder material (sealed between two glass slides) with a heating rate of 10 K/min.

**Thermogravimetric Analysis (TG) combined with a mass spectrometer (MS): TG- MS Analysis**

The thermogravimetric analysis was carried out on a Perkin Elmer Pyris 1 TGA instrument. Time-dependent temperature profiles in the range of 30 and 600 °C (20 K/min gradient) were carried out under a constant flow of $N_2$ (70 mL/min). About 2.0 mg initial sample mass was used.

Online MS measurements were carried out with a GC-Clarus 680 with an Elite-5MS glass capillary column: 30 m length, 0.25 mm diameter. MS measurements were performed on a MS Clarus SQ8C (Multiplier: 1800 V). The obtained data was processed with the TurboMass Software.

## 3. Synthesis of the Covalently Functionalized Graphene Derivatives

*Graphite intercalation compounds (GICs) with varying potassium/carbon ratios:*
**$G_{(1:4)}$** [K:C $\frac{1}{4}$ ]; **$G_{(1:8)}$** [K:C $\frac{1}{8}$ ], **$G_{(1:10)}$** [K:C $\frac{1}{10}$ ], **$G_{(1:24)}$** [K:C $\frac{1}{24}$ ], **$G_{(1:48)}$** [K:C $\frac{1}{48}$ ]

In an argon filled glove box (< 0.1 ppm oxygen; <0.1 ppm $H_2O$), spherical graphite (SGN18) and potassium – respective amounts see Table ST1 – were heated at 200 °C for 18 hours. Afterwards, the respective salt was allowed to cool to RT and isolated as a beige [K:C = 1:4], bronze [K:C = 1:8, 1:10] or black/grey [K:C = 1:24, 1:48] material (Figure S2).

| K:C | m (SGN18) [mg] | m (K) [mg] | n (GIC) [mmol] |
|---|---|---|---|
| 1:4 | 240 | 195 | 5 |
| 1:8 | 480 | 195 | 5 |
| 1:10 | 480 | 156 | 4 |
| 1:24 | 576 | 78 | 2 |
| 1:48 | 576 | 39 | 1 |

**Table TS1:** GICs obtained by the variation of the carbon/potassium ratio.

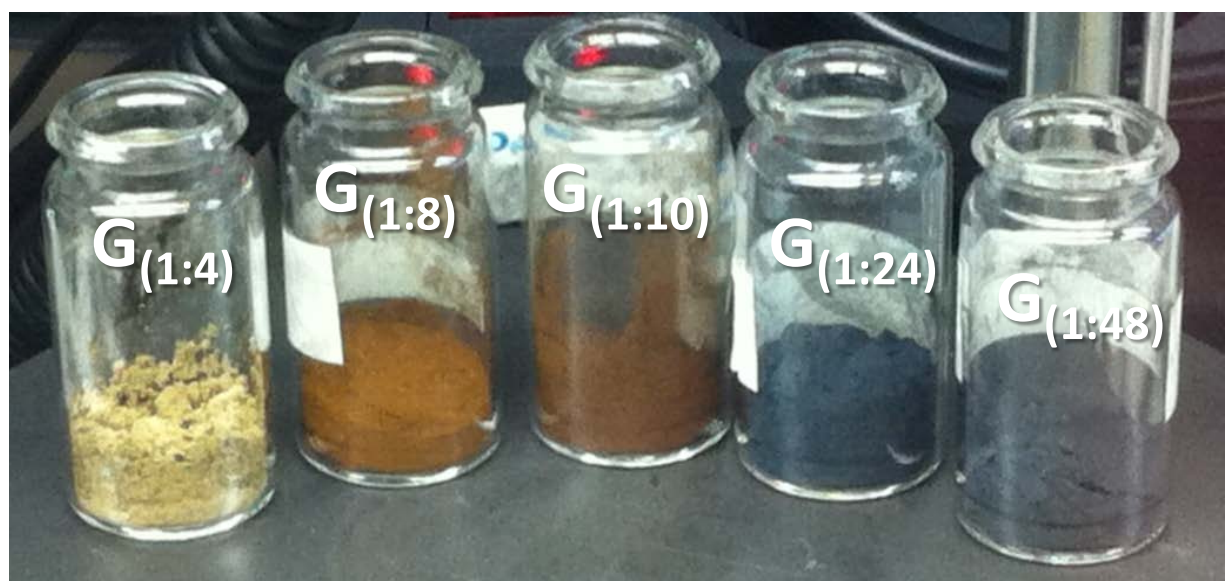

**Figure S2:** GICs with varying carbon/potassium ratio and their respective color.

*Reductive Arylation of Graphite with Phenyl Iodide – Synthesis of $G_{(1:4)}Ph$, $G_{(1:8)}Ph$, $G_{(1:10)}Ph$, $G_{(1:24)}Ph$, $G_{(1:48)}Ph$:*

1 mmol of the respective GIC (21.8 mg $G_{(1:4)}$, 16.9 mg $G_{(1:8)}$, 15.9 mg $G_{(1:10)}$, 13.6 mg $G_{(1:24)}$, 12.8 mg $G_{(1:48)}$) was dissolved in 150 mL of THF$_{abs}$ by mild tip sonication (Bandelin UW 3200, 20 J/min, puls-rate 1s, 5 min). Subsequently, 2 *eq.* of phenyl iodide (408 mg) was added and the reaction mixture was stirred at RT overnight. Afterwards, the reaction mixture was transferred from the glove box and 50 mL of water and a few drops of HCl (until pH = 7 is reached) were added to the dispersion. The reaction mixture was transferred to a separation funnel with 100 ml of cyclohexane. The phases were separated and the organic layer, containing the functionalized material, was purged three times with distilled water (200 mL each). The organic layer was filtered through a 0.2 µm reinforced cellulose membrane filter (Sartorius) and washed three times with THF, ethanol, isopropanol and water (50 mL each). The covalently functionalized material was dried at 75 °C under ambient pressure.

*Variation of the Amount of the Trapping Electrophile – Synthesis of $G_{(1:8)}Ph_{(1.6)}$, $G_{(1:8)}Ph_{(4)}$, $G_{(1:8)}Ph_{(8)}$, $G_{(1:8)}Ph_{(16)}$, $G_{(1:8)}Ph_{(40)}$, $G_{(1:8)}Ph_{(80)}$:*

1 mmol of **$G_{(1:8)}$** (16.9 mg) was dissolved in 150 mL of $THF_{abs}$ by mild tip sonication (Bandelin UW 3200, 20 J/min, puls-rate 1s, 5 min). Subsequently, the respective amout of phenyl iodide (equivalents in relation to the amount of negative charges in **$G_{(1:8)}$**: 1.6 eq.: 48 mg, 4 eq.: 106 mg, 8 eq.: 212 mg, 16 eq.: 424 mg, 40 eq.: 1060 mg; 80 eq.: 2120 mg) was added and the reaction mixture was stirred at RT overnight. Aqueous work-up and product isolation was carried out as described above.

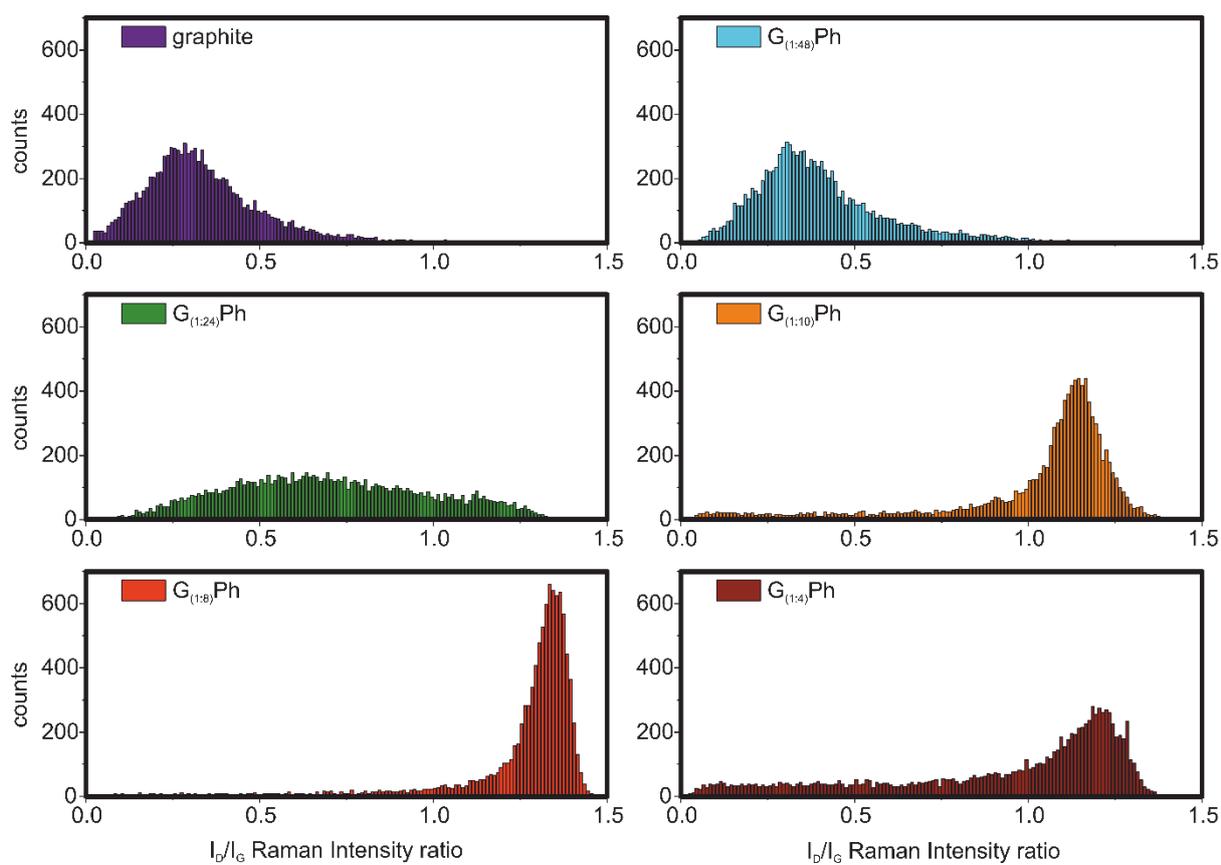

**Figure S3:** Raman determined $I_{(D/G)}$ histograms (sample area: 100 x 100 µm, 1 µm step size) of the pristine SGN18 starting material (top left) and of samples **G$_{(1:48)}$Ph** – **G$_{(1:4)}$Ph** obtained by variation of the potassium concentration.

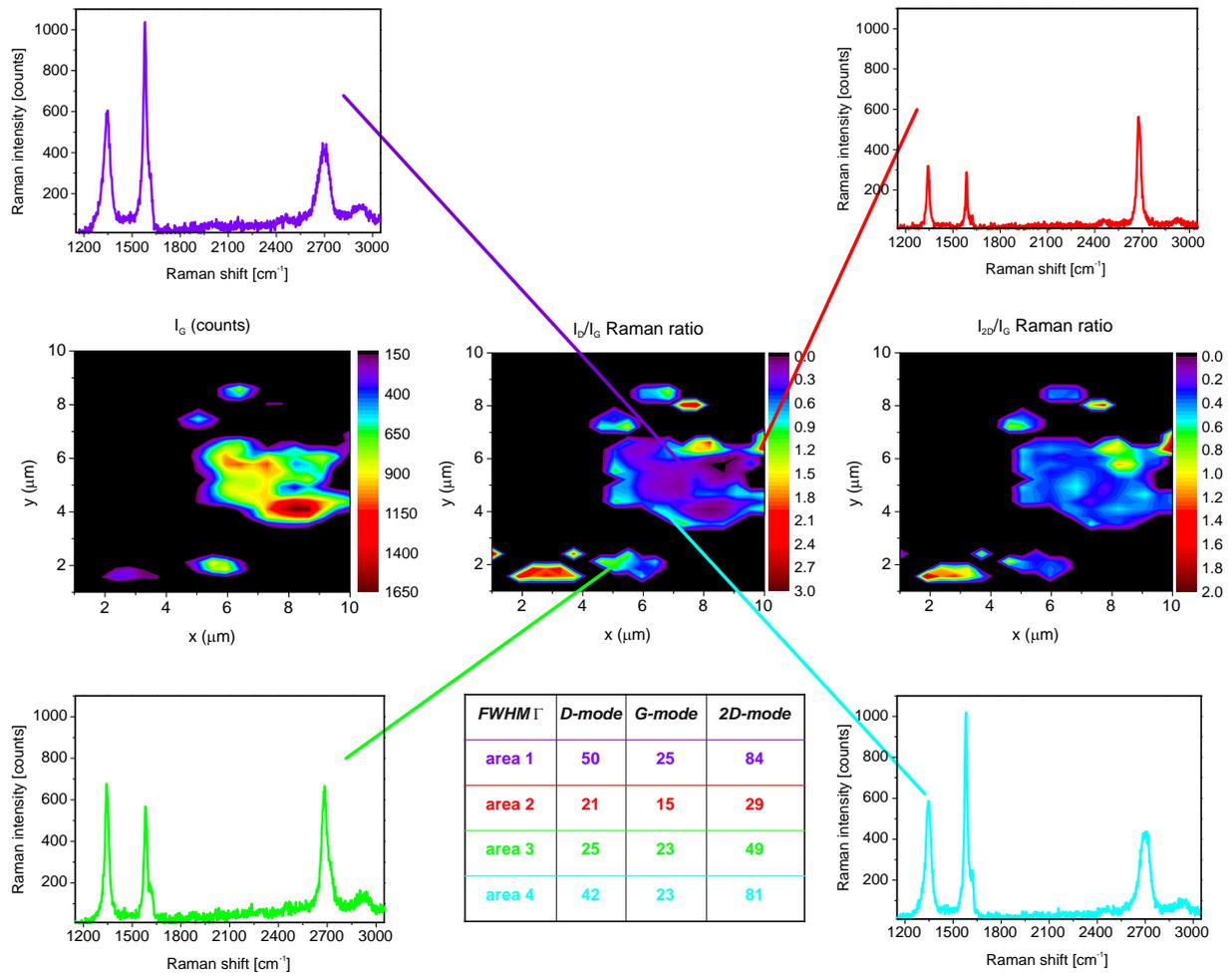

**Figure S4:** Raman analysis of deposited functionalized graphite/graphene flakes of **G$_{(1:8)}$Ph** (Si/SiO$_2$ – 300 nm oxide layer). Middle section from left to right: Special resolution plot of the G-mode intensity (identification of graphitic material), $I_{(D/G)}$ ratio (identification of the functionalized material) and $I_{(2D/G)}$ ratio (visualization of the layer character). Upper and lower section: Representative single point spectra of the respective area with fitting parameters given in the table.

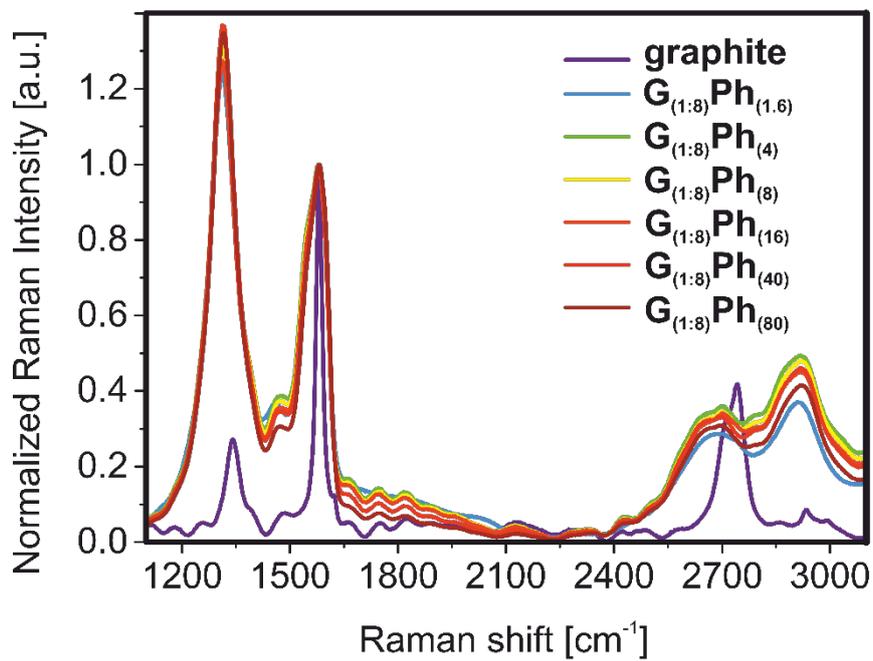
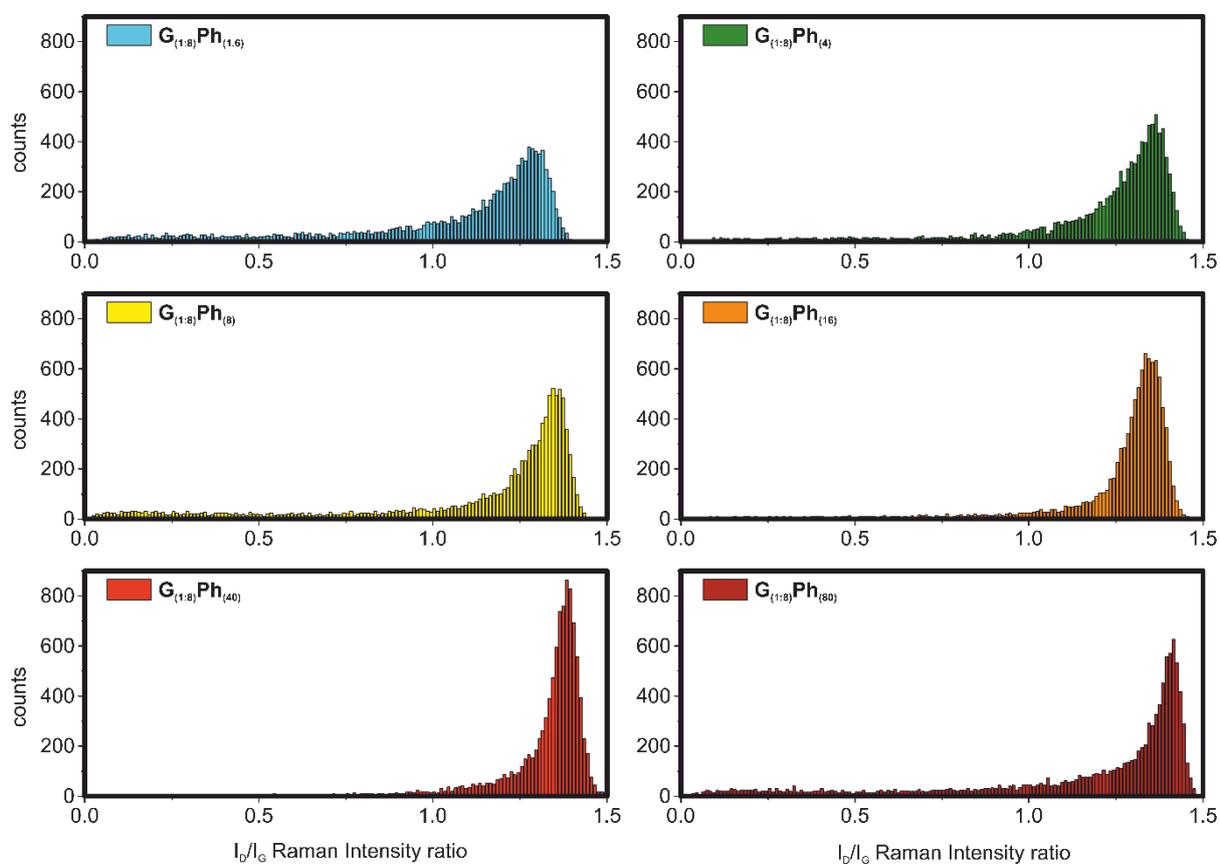

**Figure S5:** Variation of the amount of trapping electrophile in the molar excess region of negative charges present in **G$_{(1:8)}$Ph**. Top: Mean Raman spectra – normalized to G-mode intensity. Bottom: Raman determined $I_{(D/G)}$ histograms (sample area: 100 x 100 μm, 1 μm step size).

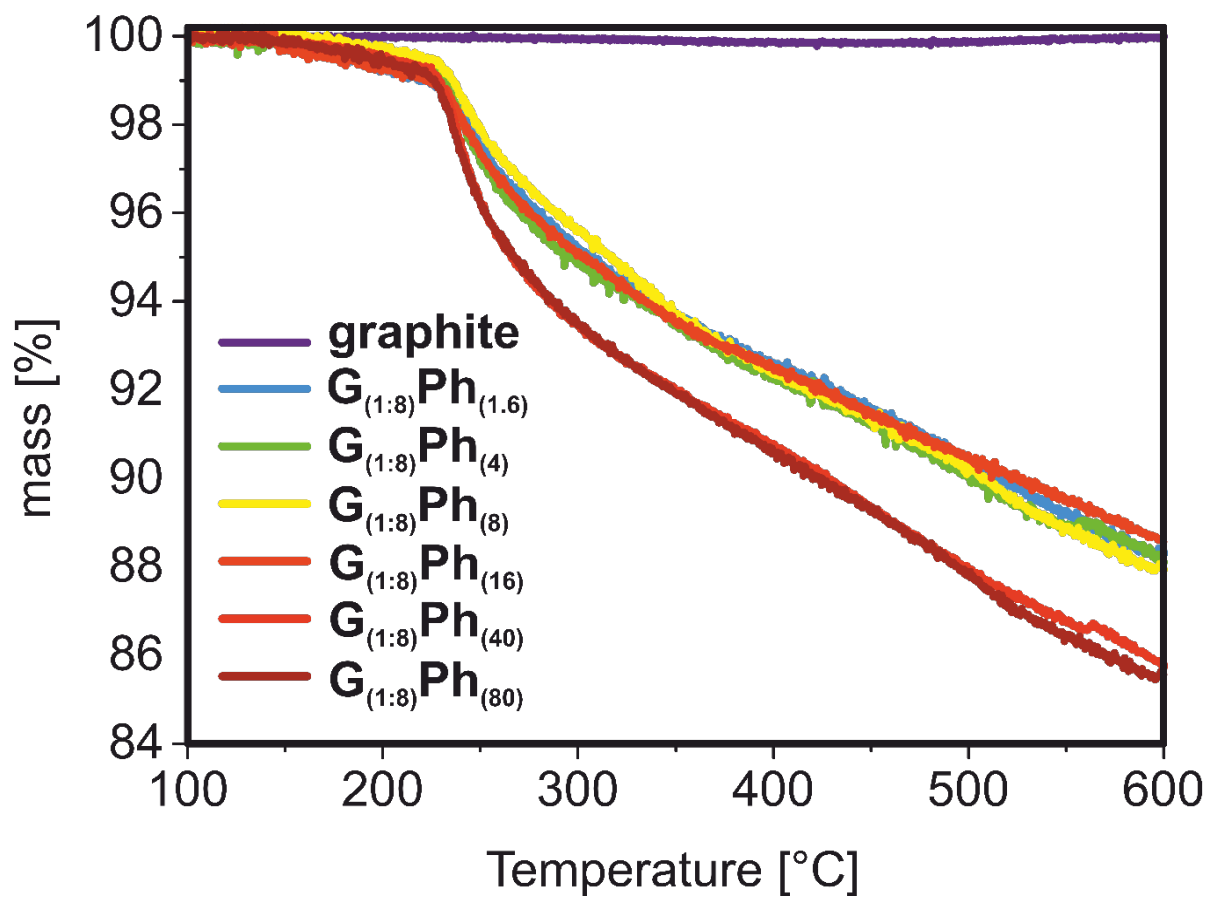

**Figure S6:** Variation of the amount of trapping electrophile in the molar excess region of negative charges present in **G$_{(1:8)}$Ph**. TG profile and mass loss in the temperature between 100 and 600 °C.

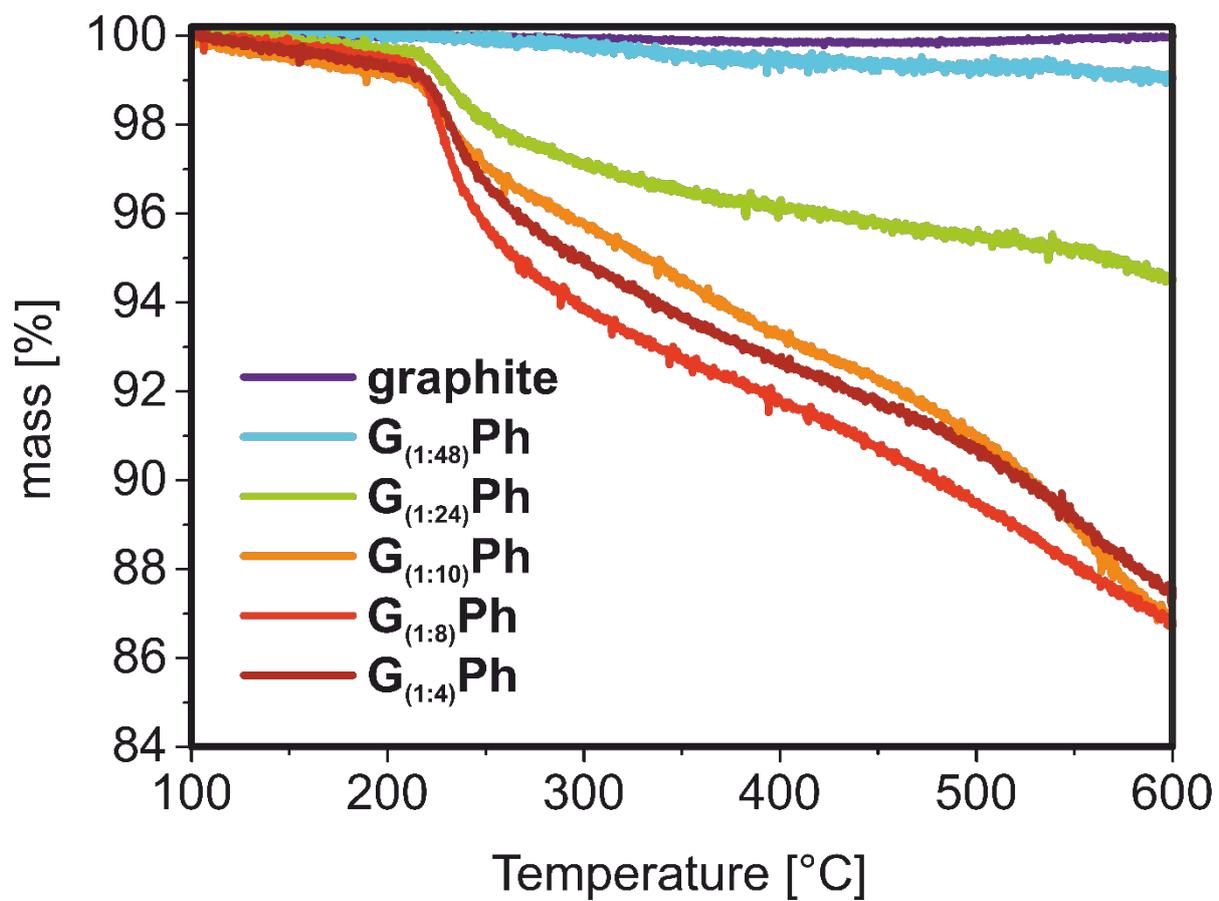

**Figure S7:** TG profile of **G$_{(1:n)}$Ph** and respective mass loss in the temperature between 100 and 600 °C.

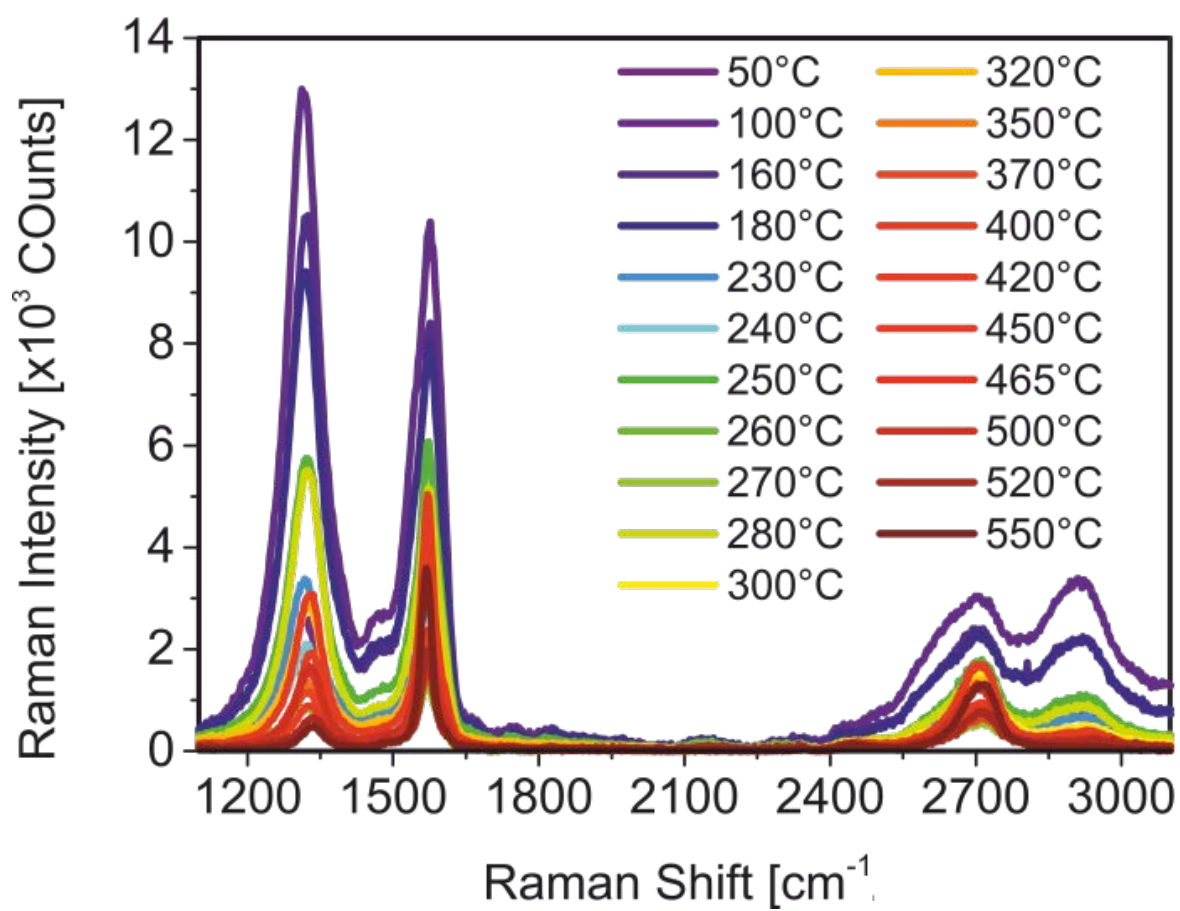

**Figure S8:** Temperature dependent Raman spectroscopy (TDRS) of **G$_{(1:8)}$Ph** in the temperature region between 50 °C and 550 °C.